\documentclass[prl,amsmath,aps,10pt,superscriptaddress,letterpaper,tightenlines, twocolumn]{revtex4}
\usepackage{bm}
\usepackage{stmaryrd}
\usepackage{amsmath}
\usepackage{amssymb}
\usepackage{graphicx}
\usepackage{textcomp}
\usepackage{calrsfs}
\usepackage{yfonts}
\usepackage{amsthm}
\usepackage{lipsum}
\newcommand{\p}{{\partial}}
\newcommand{\pt}{{\partial_t}}

\newcommand{\curl}{{\nabla\times}}
\newcommand{\haf}{{\frac{1}{2}}}
\newcommand{\intf}{{\int_0^{\infty}\,}}
\newcommand{\la}{{\langle}}
\newcommand{\ra}{{\rangle}}

\begin{document}
%%%%%%%%%%%%%%%%%%%%%%%%%%%%%%%%%%%%%%%%%%%%%%%%%%%%%%%%%%%%%%%%%%%
\title{Rotational synchronization of two non-contact nanoparticles }

\author{Vahid Ameri}
\email{vahameri@gmail.com}
\affiliation{Department of Physics, Faculty of Science, University of Hormozgan, Bandar-Abbas, Iran}
\author{M. Eghbali-Arani}
\affiliation{Department of Physics, University of Kashan, Kashan, Iran}
%%%%%%%%%%%%%%%%%%%%%%%%%%%%%%%%%%%%%%%%%%%%%%%%%%%%%%%%%%%%%%%%%%%
\begin{abstract}
Proposing a system of two rotatable nanoparticles (NPs)  in the presence of electromagnetic vacuum fluctuations, using the framework of canonical quantization, the electromagnetic and matter fields have been quantized. The non-contact frictional torque, affecting the rotation of NPs due to the presence of electromagnetic vacuum fluctuations and also by the matter field fluctuations have been derived. Considering the distance between NPs less than 100 nm in the near-field, we observe the rotations are phase locked. It has been shown that the electromagnetic vacuum fluctuations play the role of noises to break down the synchronization.  Also surprisingly, we find the frictional torque between NPs in the near-field is much bigger than the popular contact friction between them where it causes a robust synchronization in the near-field.    
\end{abstract}
%\pacs{12.20.Ds, 42.50.Lc, 03.70.+k}
\maketitle
%%%%%%%%%%%%%%%%%%%%%%%%%%%%%%%%%%%%%%%%%%%%%%%%%%%%%%%%%%%%%%%%%%%
Generally, synchronization can be treated as the coincidence on some functional of subsystems of a system due to interaction between them \cite{rosenblum1996phase}. The kinds of synchronization are widely depend on the system under consideration. In classical systems, locking the phases of the oscillators is one of the most popular coincidences of the synchronized subsystems \cite{fell2011role,rosenblum2001phase, rosenblum1996phase}.
\par Synchronization as a classical behavior has been observed in a large variety of biological, chemical, physical, and even social context \cite{strogatz2014nonlinear}. Noises play a crucial role in breaking down the synchronization. For instance, studying of synchronization in quantum systems where the quantum noises get more important, attract a lot of interests\cite{vinokur2008superinsulator,mari2013measures,goychuk2006quantum}. 
\par Quantum friction is a well-known effect affecting the motion of moving bodies \cite{dedkov2008vacuum,dedkov2012tangential,manjavacas2010thermal}. The aim of this work is to address the possibility of rotational synchronization of two NPs by considering the quantum friction as the main source for synchronization and also the noise. It is a bit confusing, however in following, we find two components of quantum friction where one is responsible for synchronization and the other can be considered as noise. The exceptional mechanical properties of NPs \cite{gieseler2012subkelvin,li2010measurement} and recently Synchronization of them \cite{gieseler2014nonlinear} have raised great interest. We investigate the rotational synchronization between two rotatable NPs considering the interactions of NPs with electromagnetic vacuum field and also with each other. The first step to this aim is to derive the quantized electromagnetic and matter fields of the system by generalizing the ideas introduced in \cite{kheirandish2014electromagnetic, huttner1992quantization} in order to study quantum friction due to the presence of the electromagnetic vacuum fluctuations and also due to the relative rotation of NPs with respect to each other. As the role of quantum friction, raised by the vacuum fluctuations, is to stop the rotation of NPs anyway, one can consider it as noise for rotational synchronization of NPs.

\par We propose a system of two NPs where one is rotating at the origin and the other one, placed a distance $ d $ from origin on the $ \mathbf{z} $ axis, is rotatable however it is initially at rest (Figure 1). Using the canonical field quantization approach, we find the explicit form of electromagnetic and medium fields in the nonrelativistic regime, then it is straightforward to derive the quantum friction torque, affecting the rotation of NPs, due to the presence in the electromagnetic vacuum field and also being next to the matter field of the other NP.
\begin{figure}
   \includegraphics[scale=0.5]{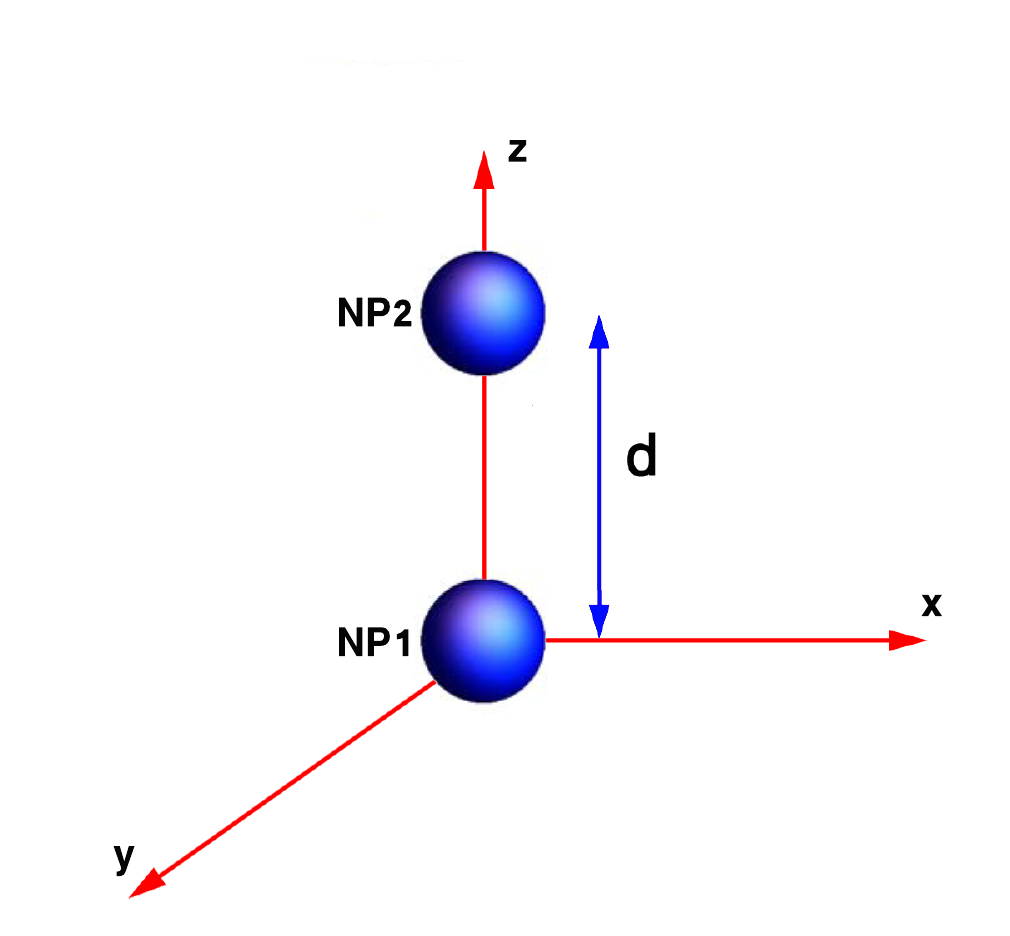}\\
  \caption{(Color online) Two NPs are located a distance $ d $ from each other on axis $ z $ .}\label{setup}
\end{figure}
\par According to the description of the system, it follows a classical regime however it will be shown that the mutual information can be considered as an order parameter for rotational synchronization of NPs.
\par We consider the following Lagrangian for a system of two spherical NPs, rotatable along their symmetric axis (z-axis), with initial angular velocity $ \omega_{0}$ for NP1 at the origin where the NP2, placed a distance $ d $ from origin on the $ \mathbf{z} $ axis, is considered to be at rest initially,  
  
\begin{eqnarray}\label{L}
\mathcal{L} &=& \haf\epsilon_0\,(\pt \mathbf{A})^2-\frac{1}{2\mu_0}(\curl\mathbf{A})^2\nonumber\\
&+&\haf\intf d\nu \,[(\pt \mathbf{X}^1+\omega_0\p_{\varphi}\mathbf{X}^1)^2-\nu^2(\mathbf{X}^1)^2]\nonumber\\
&-& \epsilon_0\intf d\nu\,f_{ij}(\nu,t)X^1_j\pt A_i\nonumber\\
&+&\epsilon_0\intf d\nu\,f_{ij}(\nu,t)X^1_j (\mathbf{v}\times\curl\mathbf{A})_i\nonumber\\
&+&\haf\intf d\nu \,[(\pt \mathbf{X}^2)^2-\nu^2(\mathbf{X}^2)^2]\nonumber\\
&-& \epsilon_0\intf d\nu\,f_{ij}(\nu,0)X^2_j\pt A_i.
\end{eqnarray}
where $ X^1 $ and $ X^2 $ are the dielectric fields describing the NP1 and NP2 respectively. The nanoparticles are considered to be in local thermodynamical equilibrium at temperature $ T $. $ f_{ij}(\nu,t) $ is the coupling tensor between  the electromagnetic vacuum field and the medium fields $ X^1 $ and $ X^2 $. As the NPs are considered to be totally similar, they should have identical coupling tensor
\begin{eqnarray}\label{C}
f(\nu,t)=\left(
  \begin{array}{ccc}
    f_{xx} (\nu) \cos(\omega_0 t) & f_{xx} (\nu) \sin(\omega_0 t) & 0 \\
    -f_{yy} (\nu) \sin(\omega_0 t) & f_{yy} (\nu) \cos(\omega_0 t) & 0 \\
    0 & 0 & f_{zz}(\nu)\\
  \end{array}
\right). \nonumber
\end{eqnarray}
One can easily derive the response functions of NPs, corresponding to set $ \omega_0=0 $, by using the coupling tensor \cite{kheirandish2014electromagnetic,ameri2015radiative},
\begin{equation}\label{kapa}
\chi^{0}_{kk}(\omega) =\epsilon_0\intf d\nu\, \frac{f^{2}_{kk} (\nu)}{\nu^2-\omega^2}.
\end{equation}
In nonrelativistic regime, we can ignore the terms containing the velocity and obtain the equations of motion,
\begin{eqnarray}\label{FE}
\mathbf{P} (\mathbf{r},\omega)&=&\mathbf{P}^{N} (\mathbf{r},\omega)+\epsilon_0 \boldsymbol{\chi} (\omega,-i\p_{\varphi})\mathbf{E}, \nonumber \\
\Bigl\{\curl\curl &-&\frac{\omega^2}{c^2}\mathbb{I}-\frac{\omega^2}{c^2}\boldsymbol{\chi}^{1}(\omega,-i \p_\varphi) -\frac{\omega^2}{c^2}\boldsymbol{\chi}^{2}(\omega) \Bigl\}\cdot\mathbf{E}\nonumber \\
&& =\mu_0\omega^2(\mathbf{P}_1^N+\mathbf{P}_2^N),
\end{eqnarray}
where $ \mathbf{P}_1^{N} $ and $ \mathbf{P}_2^{N} $ are the fluctuating or noise electric polarization components of NPs respectively \cite{kheirandish2014electromagnetic,ameri2015radiative}.
\par Using (\ref{FE}) and the dyadic Green tensor $ G_{ij} $, we find
\begin{eqnarray}\label{e}
E_i (\mathbf{r},\omega) &=& E_{0,i} (\mathbf{r},\omega)+\mu_0\omega^2\int d\mathbf{r}'\,G_{ij} (\mathbf{r},\mathbf{r}',\omega)\,\nonumber \\ && \times (P^{N}_{1,j} (\mathbf{r}',\omega)+P^{N}_{2,j} (\mathbf{r}',\omega)),
\end{eqnarray}
where the first term on the right-hand side of (\ref{e}) corresponds to the fluctuations of the electric field in electromagnetic vacuum, and the second term is the induced electric field due to the presence of NPs. One can find a proper dyadic Green's tensor $ G_{ij} $ for Eq.(\ref{e}) as
\begin{eqnarray}\label{g}
G_{ij}(\mathbf{r},\mathbf{r}',\omega)=\frac{e^{ikR}}{R^3k^2}[(k^2R^2+ikR-1)\delta_{ij}\nonumber \\
-(k^2R^2+3ikR-3)\frac{R_iR_j}{R^2}],
\end{eqnarray}
where $ \mathbf{R}=\mathbf{r}-\mathbf{r}' $ and $ k=\omega/c $.
The interparticle distance $ d $ between NPs are chosen to be much bigger than their radius, therefore, it is a good approximation to consider them as point-like particles. On the other hand, using this approximation, the components of dyadic Green's tensor $ G_{ij} $ can be simplified as
\begin{eqnarray}
G_{xx}(0,d,\omega)&=&G_{yy}(0,d\hat{z},\omega)=\frac{e^{ikd}}{d^3k^2}(k^2d^2+ikd-1), \nonumber \\
G_{zz}(0,d,\omega)&=&\frac{2e^{ikd}}{d^3k^2}(1-ikd),
\end{eqnarray}  
where all the other components are vanished.
\par The torque produced by an electric field $ E $ on a rotating particle along its rotation axis $ \hat{z} $ is 
\begin{equation}\label{m}
  \mathbf{M}=\int_V d\mathbf{r} \la \mathbf{p}(t)\times \mathbf{E}(\mathbf{r},t) \ra\cdot\mathbf{\hat z}.
\end{equation}
The frictional torque of a rotating NP in the electromagnetic vacuum  has been calculated and discussed \cite{kheirandish2014electromagnetic,manjavacas2010vacuum},
 \begin{eqnarray}\label{ms}
    \mathbf{M}_S =\frac{\hbar}{2\pi c^2}\int_0^\infty d\omega \omega^2 \mbox{Im}\,[G_{xx}(\omega) +G_{yy}(\omega) ] 
    \nonumber\\
       \,\bigg[\mbox{Im}\,[\alpha(\omega_+)][a_T(\omega_+)-a_{T_0}(\omega)]\nonumber\\
   -\mbox{Im}\,[\alpha(\omega_-)][a_T(\omega_-)-a_{T_0}(\omega)]\bigg],
 \end{eqnarray} 
 where $ \omega_\pm = \omega \pm \omega_0 $ , $ \mbox{Im}[\alpha(\omega)]= V  \mbox{Im}[\chi(\omega)] $, and $ a_T(\omega)=\coth(\hbar\omega/k_B T) $.
 The torque (\ref{ms}) generated by the components,  $\la P^N_{i}(\mathbf{r},\omega)\cdot P^N_{j}(\mathbf{r'},\omega')\ra$ and  $\la E_{0,i}(\mathbf{r},\omega)\cdot E_{0,j}(\mathbf{r'},\omega')\ra$ according to Eq.(\ref{m}), or on the other hand by the quantum friction of rotating NP due to the presence in electromagnetic vacuum field. Considering a system of two NPs, we will face another component of frictional torque which it shows the non-contact friction between NPs. Putting Eqs.(\ref{FE}) and (\ref{e}) in Eq.(\ref{m}), to derive the frictional torque on NP2, we find Eq.(\ref{ms}) and some extra components contain $\la P^N_{1,i}(\mathbf{r},\omega)\cdot P^N_{1,j}(\mathbf{r'},\omega')\ra$ which are responsible for the non-contact frictional torque between NPs.
 \par Using the fluctuation-dissipation relation
 \begin{eqnarray}
 \la P^{N}_{Bi}(\mathbf{r},\omega)P^{N\dag}_{Bj}(\mathbf{r}',\omega')\ra = 8\pi\epsilon_0 \hbar\, \mbox{Im}[\chi^{B}_{ij} (\omega)]\nonumber \\
 \times  n_T (\omega) \,\delta(\mathbf{r}-\mathbf{r}')\delta(\omega-\omega'),
 \end{eqnarray}
 we find the frictional torque $ \mathbf{M}_B $ between NPs
  \begin{eqnarray}\label{mb}
    \mathbf{M}_B =4\pi h \int_0^\infty d\omega (|G_{xx}(0,d,\omega)|^2+|G_{yy}(0,d,\omega)|^2) \nonumber\\
   \times \bigg[\big[( \mbox{Im}\alpha(\omega-\omega_{02})n_T(\omega-\omega_{02}))-( \mbox{Im}\alpha(\omega +\omega_{02}) \nonumber\\
   n_T(\omega +\omega_{02}))\big] \times [\mbox{Im}\alpha(\omega+\omega_{01})+ \mbox{Im}\alpha(\omega-\omega_{01})]
   \nonumber\\
   -\big[( \mbox{Im}\alpha(\omega-\omega_{01})n_T(\omega-\omega_{01}))-( \mbox{Im}\alpha(\omega +\omega_{01}) \nonumber\\
   n_T(\omega +\omega_{01}))\big] \times [\mbox{Im}\alpha(\omega+\omega_{02})+ \mbox{Im}\alpha(\omega-\omega_{02})] \bigg],
 \end{eqnarray} 
 where $ n_T(\omega)=1/(e^{-\beta \hbar \omega}-1) $.
\par The friction torque $ \mathbf{M}_B $ between NPs is trying to stop the rotation of NP1, while it is forcing NP2 to rotate. This interesting future of the friction possibly can make it a good candidate to synchronize the NPs. In the following, we will show that the  frictional torque $ \mathbf{M}_B $ is considerably bigger than one can expect. Domingues and et al. reported interesting results in the same case \cite{domingues2005heat}. They found that the contact conductance between two nanoparticles is smaller that the conductance for a separation distance of the order of the particles radius (in near field) where no one can expect it. We will find same results for the quantum friction between two rotating NPs.   
\par As the NP2 is initially at rest, the frictional torque $ \mathbf{M}_B $ makes it rotating and just after the beginning of the rotation, the frictional torque $ \mathbf{M}_S $, due to the presence of NP2 in the electromagnetic vacuum field, tries to stop the rotation and either the rotational synchronization. This is why we believe it plays the role of noise.
\par To have some numerical results the NPs are considered to be made of Silicon Carbide (SiC) where the dielectric function is given by the oscillator model \cite{palik1998handbook},
\begin{equation}
\varepsilon(\omega)=\varepsilon_\infty(1+\frac{\omega_L^2-\omega_T^2 }{\omega_T^2-\omega^2-i\Gamma \omega }),
\end{equation}
with $\varepsilon_\infty=6.7$, $\omega_L=1.823\times 10^{14}$, $\omega_T=1.492\times 10^{14}$, and $\Gamma = 8.954\times 10^{11}$.
The angular acceleration of NP2 could be written as
\begin{equation}\label{alf}
\alpha = \frac{\mathbf{M}}{I},
\end{equation}
where $ \mathbf{M}=\mathbf{M}_B-\mathbf{M}_S $ is the total frictional torque and $ I $ is the moment of inertia of rotating NP around its axis of symmetry. Now using $ d\omega/dt =\alpha $ where $ \alpha $ is given from Eq.(\ref{alf}) as a function of angular velocity $ \omega $, one can derive the angular velocity of NP2 as a function of time. These frictional torques also affect the angular velocity of NP1, however in this work, we considered a constant angular velocity $ \omega_1 $ for Np1.
\par Defining a proper measure of synchronization is really challenging \cite{mari2013measures,goychuk2006quantum}but not in our case. Defining the quantity 
\begin{equation}
\delta = \frac{\omega_{1}-\omega_{2}}{\omega_{1}},
\end{equation}
we get an estimate of the relative angular velocities of the NPs. For simplicity, we consider the NP1 to be at the origin rotating with constant angular velocity $ \omega_{1} $ where the NP2, placed a distance $ d $ from the origin on the $ z $ axis, rotating with angular velocity $ \omega_{2} $.
The defined quantity $ \delta $ gets values in range $ 0\leqslant \delta \leqslant 1 $. $ \delta =1 $ comes with $ \omega_{2}=0 $ which it has been considered as the initial state of the system while $ \delta=0 $ means the angular velocities of NPs are equal and on the other hand, their rotation is synchronized and also phase locked. 
\begin{figure}
   \includegraphics[scale=0.55]{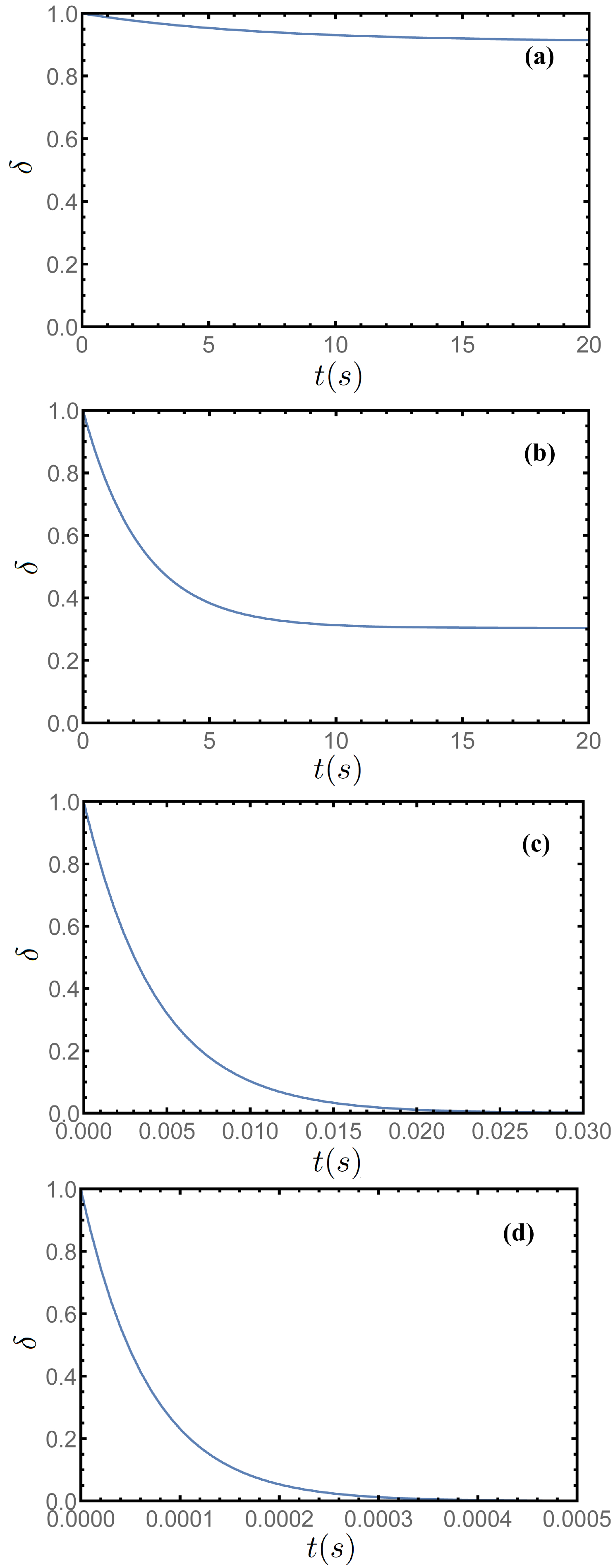}\\
  \caption{(Color online) The defined measure $ \delta $ as a function of time for different interparticle distances a) $ d=500nm $, b) $ d=200nm $, c) $ d=100nm $ and d) $ d=50nm $ where the NP1 is rotating by angular velcity $ \omega_1=10^4 rad/s $    }\label{delt}
\end{figure}
\par In figure \ref{delt} we plot $ \delta $ for different inter-particle distances as a function of time. One can easily see that, increasing the inter-particle distance is weakening the synchronization where it is not surprising because the quantum friction raised by vacuum fluctuations gets more important in compare to the quantum friction between NPs. For distances less than 100 nm in figure \ref{delt}(d) and  figure \ref{delt}(c) we have a complete phase locked synchronization. Decreasing the inter-particle distance increases the robustness of the synchronization. Mentioning the robust synchronization comes from the fact that the non-contact friction between NPs is even stronger than the popular contact frictions. It has been clearly shown that in figure \ref{delt}(c) and figure \ref{delt}(d) the angular velocities of the non-rotating NP tend to $ 10^4 rad/s $ in just $ 0.03 s $ and $ 0.0005 s $ respectively, which it is almost impossible with popular contact friction of particles.
\par  Increasing the inter-particle distances in figure \ref{delt}(a) and figure \ref{delt}(b), we lose the phase locked rotation of NPs, however, we may have a resonant angular velocity entrainment condition, as the $ \delta $ tends to a finite constant value.
\par Recently, authors introduced the mutual information as an order parameter for quantum synchronization which it can qualify the synchronization on a large variety of systems from semi-classical continuous-variable systems to the deep quantum ones \cite{ameri2015mutual}. Mutual information of system $AB$ composed of two subsystem $A$ and $B$ is defined as
\begin{equation}
I=S_A+S_B-S_{AB}
\end{equation} 
where $ S_A $ and $ S_B $ are the entropies of subsystem $ A $ and $ B $ respectively and $ S_{AB} $ shows the total entropy of the system. Mutual information qualifies how much the knowledge of the subsystem $ A $ gives information about the subsystem $ B $ \cite{cover1991entropy}. When the NPs get closer, heat transfer between them increases and so the flow of entropy between them \cite{pendry1999radiative,pendry1983quantum}. One can say, increasing the flow of entropy, decreases the total entropy of the system. On the other hand, increasing the flow of entropy between the subsystems increases the mutual information of the system. As mentioned before, mutual information has been introduced as an order parameter for quantum synchronization \cite{ameri2015mutual}, where increasing in the mutual information considered as a signal of the presence of the quantum synchronization. Interestingly here, figure \ref{delt} shows rotational synchronization of NPs as they get closer where the mutual information of NPs increases.
 \par In conclusion, we have demonstrated the possibility of rotational synchronization of two neutral NPs through the electromagnetic matter field fluctuations or on the other hand the quantum friction between them. Surprisingly, the quantum friction between NPs in the near-field was much bigger than our expectation and even bigger than the normal contact friction between NPs. Same results on the heat transfer between two NPs had been reported previously. We show that the electromagnetic vacuum field fluctuations play the role of noises and are weakening the rotational synchronization of NPs. We emphasize that, there is a connection between the rotational synchronization of NPs in this work and the mutual information, reported previously as an order parameter of quantum synchronization.

\bibliography{sci}
\bibliographystyle{iopart-num}
\end{document}